\documentclass[a4paper,11pt]{article}
\usepackage{pos}

\usepackage{mathtools}
\usepackage{amsfonts}
\usepackage{mathrsfs}
\usepackage{bbm}
\usepackage{slashed}
\usepackage{amsmath}
\usepackage{bm}
\usepackage{bbold}
\usepackage{booktabs}
\usepackage[nameinlink]{cleveref}

\graphicspath{{./figures/}}

\makeatletter
\g@addto@macro\normalsize{%
  \setlength\abovedisplayskip{9pt plus 2pt minus 2pt}%
  \setlength\belowdisplayskip{9pt plus 2pt minus 2pt}%
  \setlength\abovedisplayshortskip{2pt plus 2pt}%
  \setlength\belowdisplayshortskip{4pt plus 2pt minus 2pt}}
\makeatother
\newcommand{\MeV}{\,\text{MeV}}
\newcommand{\fm}{\,\text{fm}}
\newcommand{\rmi}{\mathrm{i}}

\title{Density functional theory of renormalization group in nuclear matter}

\author[a,b]{Yong-rui Chen}
\author[a,c,d]{Yang-yang Tan}
\author*[a]{Wei-jie Fu}

\affiliation[a]{School of Physics, Dalian University of Technology,\\ Dalian, 116024, P.R. China}
\affiliation[b]{Institut f\"ur Theoretische Physik, Justus-Liebig-Universit\"at Gie\ss en,\\ 35392 Gie\ss en, Germany}
\affiliation[c]{Institute for Physics of Intelligence, Graduate School of Science, The University of Tokyo, Bunkyo-ku, Tokyo 113-0033, Japan}
\affiliation[d]{RIKEN Center for Interdisciplinary Theoretical and Mathematical Sciences (iTHEMS), Wako, Saitama 351-0198, Japan}

\emailAdd{yong-rui.chen@uni-giessen.de}
\emailAdd{yangyang-tan@foxmail.com}
\emailAdd{wjfu@dlut.edu.cn}

\abstract{
We present the density functional renormalization group (density-fRG), in which density fluctuations are separated from the quantum fluctuations in the vacuum and are integrated out successively within the functional renormalization group approach. This allows us to quantify medium effects and to study the physics of high baryon densities. The method is applied to the nucleon-meson effective field theory, also known as the Walecka model, where the density dependence of the nucleon-meson Yukawa couplings is generated self-consistently by their flow equations. It is found that both the attractive and the repulsive nucleon-meson interactions are screened by the dense medium. This results in an equation of state of symmetric nuclear matter that is stiffer than in the relativistic mean-field approximation for $\rho_0\lesssim\rho\lesssim 2.5\,\rho_0$, and becomes softer for $\rho\gtrsim 2.5\,\rho_0$, where $\rho_0$ is the saturation density. Furthermore, a new phenomenon called the locking of the Fermi surface is found: the effective energy of the quasi-nucleon stays close to the Fermi surface, while both are running with the renormalization group scale.
}

\FullConference{%
  The 13th International Conference on the Exact Renormalization Group (ERG2026)\\
  1--5 September 2026\\
  University of Sussex, Brighton, United Kingdom
}

\begin{document}
\maketitle

%%%%%%%%%%%%%%%%%%%%%%%%%%%%%%%%%%%%%%%%%%%%%%%%%%%%%%%%%%%%%%%%%%%%%%%%%%%%%%%
\section{Introduction}
\label{sec:int}

Strongly interacting matter at high baryon densities has attracted increasing attention in recent years. Heavy-ion experiments \cite{STAR:2020tga,STAR:2025zdq} and theoretical studies \cite{Fu:2019hdw, Gao:2020fbl, Gunkel:2021oya} indicate that the critical end point of the QCD phase diagram is more likely located at large baryon chemical potential $\mu_B$, see \cite{Fu:2022gou} for a review. Moreover, the equation of state (EoS) at high densities is a key input for neutron stars and their gravitational-wave signatures \cite{LIGOScientific:2017vwq}. With the fixed-target program at RHIC reaching $\mu_B\sim 750\MeV$ \cite{STAR:2025zdq} and new facilities such as CBM \cite{Friman:2011zz}, CEE \cite{Hu:2023niz}, HIAF \cite{Yang:2013yeb} and NICA \cite{Sorin:2011zz}, reliable theoretical computations at large $\mu_B$ are called for. Lattice QCD simulations are restricted to $\mu_B/T\lesssim 2\sim 3$ by the sign problem \cite{Borsanyi:2020fev}, while first-principles QCD calculations within the functional renormalization group (fRG) reliably reach $\mu_B/T\sim 4$ \cite{Fu:2019hdw}; see \cite{Dupuis:2020fhh, Fu:2022gou,Fischer:2026uni} for reviews.

In the fRG, quantum, thermal and density fluctuations are taken into account successively with the evolution of the renormalization group (RG) scale $k$. In the spirit of density functional theory, we separate the density fluctuations from the quantum fluctuations in the vacuum and only compute the former with the RG flow, which simplifies the calculations at high densities considerably. We call this method the density functional theory of renormalization group, or density functional renormalization group (density-fRG), and summarize here its first application to the nucleon-meson effective field theory (EFT), i.e., the Walecka model \cite{Walecka:1974qa}; see \cite{Chen:2025ijh} for details.

%%%%%%%%%%%%%%%%%%%%%%%%%%%%%%%%%%%%%%%%%%%%%%%%%%%%%%%%%%%%%%%%%%%%%%%%%%%%%%%
\section{Density-fRG for the nucleon-meson effective field theory}
\label{sec:method}

The effective action of the nucleon-meson EFT reads, in Euclidean spacetime,
\begin{align}
  &\Gamma_{k}[\Phi]=\int_x \bigg\{ \bar{\psi}_N \Big[\gamma_\mu \partial_\mu+m_N-\gamma_0\hat\mu_N-h_{\sigma,k}\,\sigma\, T^{0}
  +\rmi\, h_{\omega,k}\,\gamma_\mu \omega_\mu T^{0}+\rmi\, h_{\rho,k}\,\gamma_\mu \bm{\rho}_\mu\cdot \bm{T}\Big]\psi_N\nonumber\\[1ex]
  &+\frac{1}{2}(\partial_\mu \sigma)^2+U(\sigma)
  +\frac{1}{4}\omega_{\mu \nu}\omega_{\mu \nu}+\frac{1}{2}m_\omega^2\omega_\mu \omega_\mu
  +\frac{1}{4}\bm{\rho}_{\mu \nu}\cdot\bm{\rho}_{\mu \nu}+\frac{1}{2}m_\rho^2\bm{\rho}_\mu \cdot\bm{\rho}_\mu+V_k(\sigma,\omega,\rho)\bigg\}\,,
  \label{eq:action}
\end{align}
with the nucleon doublet $\psi_N=(\psi_p,\psi_n)^{\intercal}$, the isoscalar-scalar $\sigma$, isoscalar-vector $\omega_\mu$ and isovector-vector $\bm{\rho}_\mu$ mesons, the isospin generators $T^0=\mathbb{1}/2$ and $T^i=\tau^i/2$ with the Pauli matrices $\tau^i$, $\hat\mu_N=\mathrm{diag}(\mu_p,\mu_n)$, and the $\sigma$ potential $U(\sigma)=m_\sigma^2 \sigma^2/2+\lambda_3\sigma^3/3!+\lambda_4\sigma^4/4!$. The vacuum quantities $m_N$, $m_{\sigma}$, $m_\omega$, $m_\rho$ and $\lambda_{3,4}$ are $k$-independent, whereas the Yukawa couplings $h_{\sigma,k}$, $h_{\omega,k}$ and $h_{\rho,k}$ run with $k$. The effective potential $V_k(\sigma,\omega,\rho)$ arises solely from density fluctuations and thus vanishes in the vacuum, $V_{k\to\infty}=0$, while the full medium effects are encoded at $k\to 0$.
%%%%%%%%%%%%%%%%%%%%%%%%%%%%%
\begin{figure}[t]
  \centering
  \includegraphics[width=0.46\textwidth]{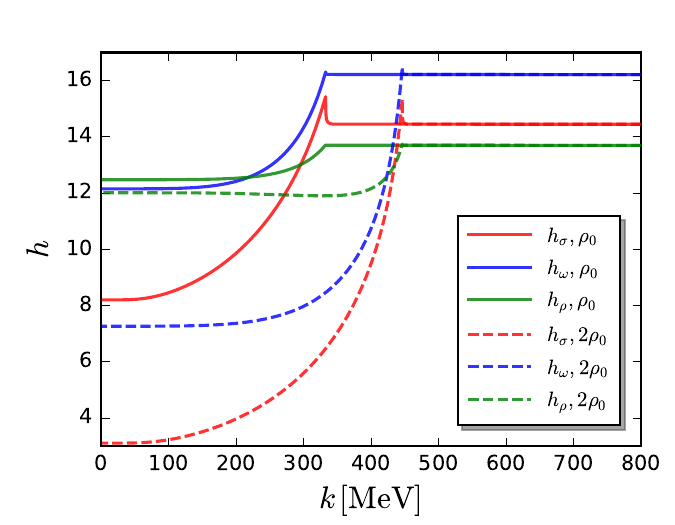}\hfill
  \includegraphics[width=0.46\textwidth]{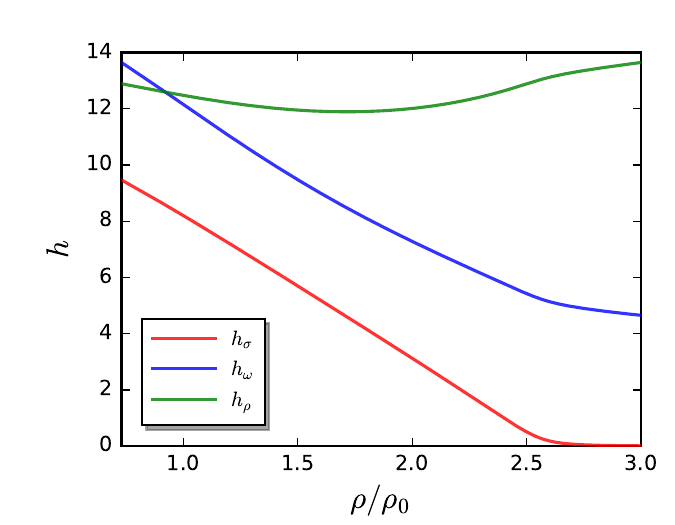}
  \caption{Left panel: Yukawa couplings as functions of the RG scale $k$ at the baryon densities $\rho_0$ and $2\rho_0$ in symmetric nuclear matter. Right panel: Infrared value of Yukawa couplings, $h_{k=0}$, as functions of the baryon density. Both panels are obtained with $m_\sigma=500\MeV$.}
  \label{fig:h}
\end{figure}
%%%%%%%%%%%%%%%%%%%%%%%%%%%%%

The RG flow of \labelcref{eq:action} is governed by the Wetterich equation~\cite{Wetterich:1992yh}, which we evaluate with $3d$ flat regulators~\cite{Litim:2000ci, Litim:2001up} on the equations of motion (EoM), where only $\sigma$ and the temporal components $\omega_0$ and $\rho_{03}$ (isospin index 3) of the vector mesons, which are purely imaginary, have nonvanishing expectation values. The flow of $V_k$ is given by the nucleon loop with its vacuum part ($T=\mu=0$) subtracted, so that only density fluctuations remain; the meson loop does not contribute, as mesons carry no chemical potential. Integrating down to $k=0$ yields the thermodynamic potential density
\begin{align}
  \Omega=-\frac{1}{3 \pi^2}\int_0^\infty \mathrm{d}k\, \frac{k^4}{E_k}\sum_{i=p,n}\,\sum_{\pm}n_F(E_k\pm\mu_i^*)
  +U(\sigma)-\frac{1}{2}m_\omega^2 (\rmi\omega_0)^2 -\frac{1}{2}m_\rho^2(\rmi \rho_{03})^2\,,
  \label{eq:Omega}
\end{align}
with $E_k=(k^2+m_N^{*2})^{1/2}$, $n_F(x)=1/(e^{x/T}+1)$, the effective nucleon mass $m_N^*= m_N-h_{\sigma,k}\sigma/2$, and the effective chemical potentials $\mu_{p,n}^* =\mu_{p,n}-(h_{\omega,k}\,\rmi\omega_0\pm h_{\rho,k}\, \rmi\rho_{03})/2$, which depend on $k$ through the running Yukawa couplings. The fields follow from the EoM $\partial \Omega/\partial \sigma=\partial \Omega/\partial \omega_0=\partial \Omega/\partial \rho_{03}=0$. For constant Yukawa couplings, \labelcref{eq:Omega} is precisely the relativistic mean-field (RMF) result, see, e.g.,~\cite{Bai:2021wrh}, with $k$ playing the role of the momentum for the $3d$ flat regulator. Hence, the density-fRG goes beyond RMF through the running Yukawa couplings, which encode the density dependence of the interactions self-consistently.

The flows of the Yukawa couplings follow from those of the nucleon-meson vertices, which include diagrams with $\sigma$, $\omega$ and $\rho$ exchange and with the $\sigma$ self-interactions~\cite{Chen:2025ijh}, projected onto vanishing meson momentum and nucleon momenta $p_0=(\pi T,\bm{0})$ at the lowest Matsubara frequency,
\begin{align}
  \partial_t h_{\sigma,k}&=-\frac{1}{2}\,\mathrm{Re}\Big[\mathrm{Tr}\Big(T^0\,\partial_t \Gamma^{(3)}_{\bar \psi \psi \sigma}(p_0)\Big)\Big]\,,\nonumber\\
  \partial_t h_{\omega,k}&=-\frac{1}{6}\,\mathrm{Re}\Big[\mathrm{Tr}\Big(T^0\, \mathrm{i} \vec{\gamma}_\mu\,\partial_t \big(\Gamma^{(3)}_{\bar \psi \psi \omega}\big)_\mu(p_0)\Big)\Big]\,,\nonumber\\
  \partial_t h_{\rho,k}&=-\frac{1}{6}\,\mathrm{Re}\Big[\mathrm{Tr}\Big(T^3\, \mathrm{i} \vec{\gamma}_\mu\,\partial_t \big(\Gamma^{(3)}_{\bar \psi \psi \rho}\big)_\mu^{i=3}(p_0)\Big)\Big]\,,
  \label{eq:dth}
\end{align}
with the RG time $t=\ln(k/\Lambda)$; the trace runs over Dirac and isospin space, and the real part averages over $\pm p_0$. With $\vec{\gamma}_\mu=(1-\delta_{\mu 0})\gamma_\mu$, the vector vertices are projected onto their spatial components, neglecting the difference to the temporal ones. The meson propagators are kept at their vacuum form, and the vacuum parts of the threshold functions are subtracted as for $V_k$. The Silver Blaze property \cite{Cohen:2003kd}, violated since the external nucleon frequency is fixed rather than summed over, is partially recovered by multiplying the threshold functions with $\sum_\pm n_F(E_k\pm\mu^*)$, a factor emerging from the frequency summation in two-loop fRG calculations \cite{Fu:2016tey}. At $T=0$ it reduces to $\Theta(\mu^*-E_k)$, i.e., the couplings only flow below the Fermi surface.
%%%%%%%%%%%%%%%%%%%%%%%%%%%%%%%%%%%%%%%%%%%%%%%%%%%%%%%%%%%%%%%%%%%%%%%%%%%%%%%
\section{Numerical results}
\label{sec:num}

We use $m_N=938\MeV$, $m_\omega=782\MeV$, $m_\rho=775\MeV$ \cite{ParticleDataGroup:2024cfk}, and $m_\sigma=500$ or $600\MeV$ for the poorly constrained $\sigma$ mass. The vacuum values of the Yukawa couplings and $\lambda_{3,4}$ are fitted to symmetric nuclear matter at saturation: $\rho_0=0.16\fm^{-3}$, the binding energy per nucleon $E/A=-16\MeV$, the incompressibility $K=215\MeV$ and the symmetry energy $E_s=32\MeV$ \cite{Chen:2025ijh}. RMF results with constant couplings fitted likewise are shown for comparison. We use $T=3\MeV$.

The left panel of \Cref{fig:h} shows the Yukawa couplings as functions of $k$ at $\rho=\rho_0$ and $2\rho_0$. For large $k$ there are no density fluctuations and the couplings keep their vacuum values. Below a threshold scale, which increases with the density, density fluctuations set in and, apart from a small initial rise of $h_{\sigma,k}$, the couplings decrease towards the infrared: both the attractive nucleon-$\sigma$ and the repulsive nucleon-$\omega$ interactions are screened by the medium, where the vector-meson exchange in the flow of $h_{\sigma,k}$ plays a pivotal role in screening the $\sigma$ channel \cite{Chen:2025ijh}. At $k=0$ (right panel), $h_{\sigma}$ and $h_{\omega}$ decrease with the density, and $h_{\sigma}$ almost vanishes for $\rho\gtrsim 2.5\,\rho_0$, whereas $h_\rho$ varies mildly.

%%%%%%%%%%%%%%%%%%%%%%%%%%%%%
\begin{figure}[t]
  \centering
  \includegraphics[width=0.46\textwidth]{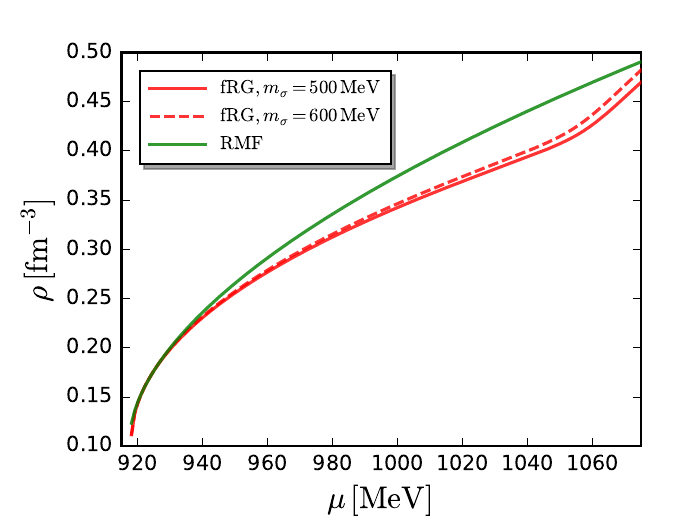}\hfill
  \includegraphics[width=0.46\textwidth]{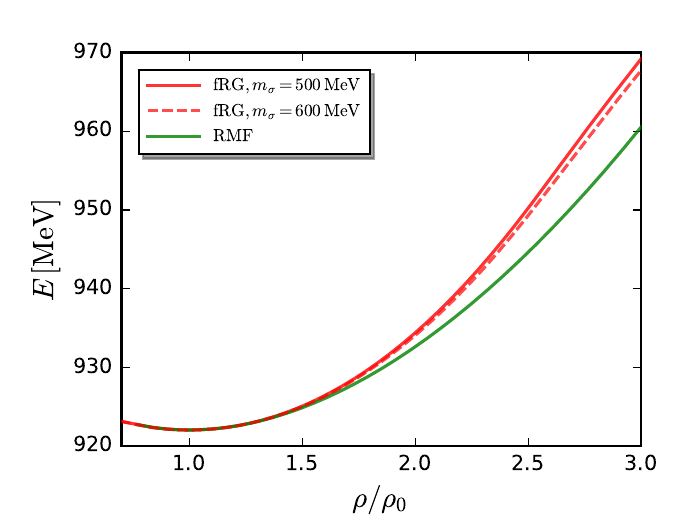}\\[0.5ex]
  \includegraphics[width=0.46\textwidth]{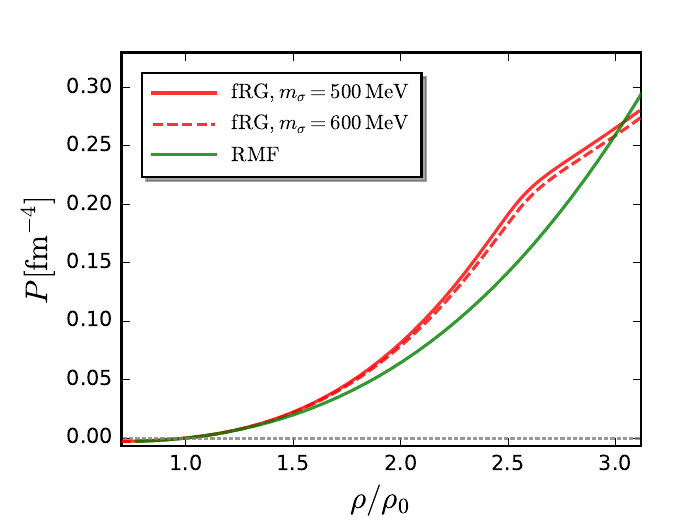}\hfill
  \includegraphics[width=0.46\textwidth]{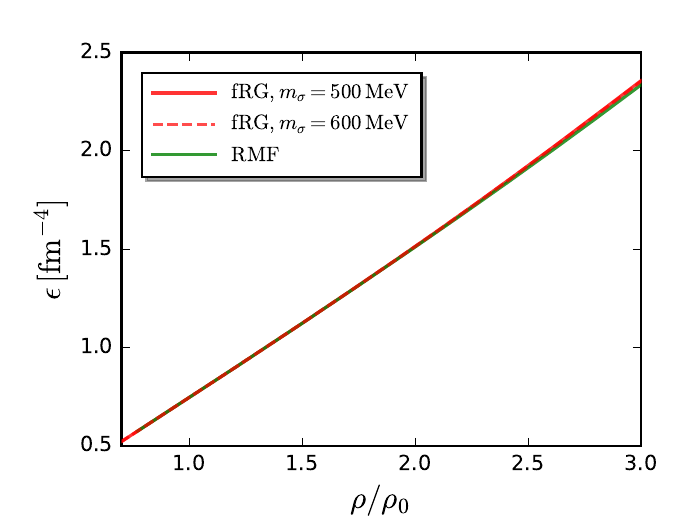}
  \caption{Equation of state of symmetric nuclear matter: nucleon density as a function of the nucleon chemical potential $\mu=\mu_p=\mu_n$ (top left), and energy per nucleon (top right), pressure (bottom left) and energy density (bottom right) as functions of the nucleon density. The fRG results with $m_\sigma=500$ and $600\MeV$ are compared with RMF.}
  \label{fig:eos}
\end{figure}
%%%%%%%%%%%%%%%%%%%%%%%%%%%%%

In contrast to RMF, the running couplings render $m_N^*$ and $\mu^*$ $k$-dependent. The couplings only flow for $E_k<\mu^*$, i.e., below the Fermi surface. There, the screening increases $m_N^*$ and pushes $E_k$ towards $\mu^*$, but not beyond, since the flows would cease otherwise. Consequently, $E_k$ stays close to the running $\mu^*$ in a sizable range of $k$ below the threshold scale. We call this novel phenomenon, absent in RMF, the locking of the Fermi surface \cite{Chen:2025ijh}. Moreover, contrary to RMF, the effective nucleon mass at $k=0$ increases with the density and approaches its vacuum value at $\rho\approx 2.5\,\rho_0$.

The pressure $p=-\Omega$, the nucleon density $\rho=\rho_p+\rho_n$ with $\rho_{p,n}=-\partial\Omega/\partial\mu_{p,n}$, the energy density $\epsilon=Ts+\mu_p\rho_p+\mu_n\rho_n-p$ with $s=-\partial\Omega/\partial T$, and the energy per nucleon $E=\epsilon/\rho$ are shown in \Cref{fig:eos}. Both fRG parameter sets lead to very similar results. The pressure in the fRG increases faster than in RMF for $\rho\lesssim 2.5\,\rho_0$, i.e., the EoS is stiffer, owing to the screening of the nucleon-meson interactions. For $\rho\gtrsim 2.5\,\rho_0$ the screening is almost completed, cf.\ $h_{\sigma,k=0}$ in \Cref{fig:h}, and the increase of the pressure slows down, i.e., the EoS becomes softer. Accordingly, the fRG density rises more slowly with $\mu$ than the RMF one and steepens around $2.5\,\rho_0$, and the fRG energy per nucleon grows faster above saturation, while the energy densities are barely distinguishable.

The reliability of the nucleon-meson EFT at high densities can be crudely estimated by the ratio $R=2r_N\rho^{1/3}$ of the nucleon diameter to the average inter-nucleon distance. With the proton charge radius $r_N=0.84\sim 0.88\fm$ \cite{Lee:2015jqa, Bezginov:2019mdi}, one finds $R=0.91\sim 0.96$, $1.04\sim 1.09$ and $1.24\sim 1.30$ at $\rho=\rho_0$, $1.5\,\rho_0$ and $2.5\,\rho_0$, respectively. Hence, nucleons begin to overlap at $\rho\approx 1.5\,\rho_0$, and their substructure, absent in the EFT, should matter for $\rho\gtrsim(2.5\sim 3)\,\rho_0$.

%%%%%%%%%%%%%%%%%%%%%%%%%%%%%%%%%%%%%%%%%%%%%%%%%%%%%%%%%%%%%%%%%%%%%%%%%%%%%%%
\section{Conclusions and outlook}
\label{sec:conclusion}

We have proposed the density-fRG and applied it to study the properties and the EoS of nuclear matter at high densities. The underlying separation is supported by the fact that the two dominate in different regimes of the RG scale: density fluctuations mainly for $k\lesssim 400\sim 500\MeV$, quantum fluctuations above, see, e.g., \cite{Fu:2019hdw}. Applied to the Walecka model, the density-fRG shows that the medium screens both the attractive nucleon-$\sigma$ and the repulsive nucleon-$\omega$ interactions. This leads to the locking of the Fermi surface and to an EoS of symmetric nuclear matter that is stiffer than in RMF for $\rho_0\lesssim\rho\lesssim 2.5\,\rho_0$ and becomes softer for $\rho\gtrsim 2.5\,\rho_0$, where, however, the reliability of the nucleon-meson EFT is gradually lost. The symmetry energy at high densities is under investigation. Further extensions include pion fluctuations, absent in mean field but sizable due to the small pion mass, hyperons, and transport-model applications to heavy-ion collisions. Moreover, summing the Matsubara frequencies of the external nucleon line in the Yukawa flows yields two-loop flows, cf.~\cite{Fu:2016tey}, which restore the Silver Blaze property and generate new distribution functions such as $n_F(E_N\pm E_\phi-\mu)$ with nucleon and meson energies $E_{N,\phi}$ \cite{Chen:2026tlf}.

%%%%%%%%%%%%%%%%%%%%%%%%%%%%%%%%%%%%%%%%%%%%%%%%%%%%%%%%%%%%%%%%%%%%%%%%%%%%%%%
\acknowledgments
We thank Jan M. Pawlowski for discussions and comments. We also thank the members of the fQCD collaboration \cite{fQCD} for collaborations on related projects. This work is supported by the National Natural Science Foundation of China under Contract Nos.\ 12447102, 12175030. \mbox{Y.-r.} Chen is supported by the Alexander von Humboldt Foundation through a Humboldt Research Fellowship for postdocs.

\bibliographystyle{JHEP}
\bibliography{refs}

\end{document}